\documentclass{article}

\usepackage{arxiv}

\usepackage[utf8]{inputenc} 
\usepackage[T1]{fontenc}    
\usepackage{hyperref}       
\usepackage{url}            
\usepackage{booktabs}       
\usepackage{amsfonts}       
\usepackage{nicefrac}       
\usepackage{microtype}      
\usepackage{lipsum}
\usepackage{natbib}

\usepackage{changepage}
\usepackage{textcomp,marvosym}
\usepackage{fixltx2e}
\usepackage{amsmath,amssymb}
\usepackage{cite}
\usepackage{nameref,hyperref}
\usepackage[right]{lineno}
\usepackage{soul}
\usepackage{microtype}
\usepackage{color}
\usepackage{lastpage,fancyhdr,graphicx}
\usepackage{epstopdf}

\title{Eco-evolutionary Model of Rapid Phenotypic Diversification in Species-Rich Communities.}

\author{
  Paula Villa Mart{\'\i}n \\
\thanks{Current affiliation: Biological Complexity Unit,
  Okinawa Institute of Science and Technology Graduate University,
  Onna, Okinawa 904-0495, Japan.}
  Departamento de Electromagnetismo y F{\'i}sica de la Materia \\
  and Instituto Carlos I de F{\'i}sica Te{\'o}rica y
  Computacional,
  Universidad de Granada, Granada, Spain.\\
   \And
  Jorge Hidalgo \\
  Departamento de Electromagnetismo y F{\'i}sica de la Materia \\
  and Instituto Carlos I de F{\'i}sica Te{\'o}rica y
  Computacional,
  Universidad de Granada, Granada, Spain.\\
  Dipartimento di Fisica 'G.Galilei' and CNISM,
  INFN, Universit\`a di Padova, Padova, Italy.
  \And
  Rafael Rubio de Casas \\
  Estaci\'on Experimental de Zonas \'Aridas, EEZA-CSIC, Almer\'ia, Spain.\\
  UMR 5175 Centre d’Ecologie Fonctionnelle et Evolutive (CNRS), Montpellier, France. \\
  Departamento de Ecolog\'ia, Universidad de Granada, Granada, Spain.\\
  \AND
  Miguel A. Mu\~noz \\
  Departamento de Electromagnetismo y F{\'i}sica de la Materia \\
  and Instituto Carlos I de F{\'i}sica Te{\'o}rica y
  Computacional,
  Universidad de Granada, Granada, Spain.\\
  \texttt{mamunoz@onsager.ugr.es}
}

\begin{document}
\maketitle

\begin{abstract}
Evolutionary and ecosystem dynamics are often treated as different
processes --operating at separate timescales-- even if evidence
reveals that rapid evolutionary changes can feed back into ecological
interactions. A recent long-term field experiment has explicitly
  shown that communities of competing plant species can experience
  very fast phenotypic diversification, and that this gives rise to
  enhanced complementarity in resource exploitation and to enlarged
  ecosystem-level productivity. Here, we build on progress made in
  recent years in the integration of eco-evolutionary dynamics, and
  present a computational approach aimed at describing these empirical
  findings in detail.  In particular we model a community of organisms
  of different but similar species evolving in time through
  mechanisms of birth, competition, sexual reproduction, descent with
  modification, and death. Based on simple rules, this model provides
  a rationalization for the emergence of rapid phenotypic
  diversification in species-rich communities.
  Furthermore, it also leads to non-trivial predictions about
  long-term phenotypic change and ecological interactions. Our results
  illustrate that the presence of highly specialized, non-competing
  species leads to very stable communities and reveals that
  phenotypically equivalent species occupying the same niche may
  emerge and coexist for very long times. Thus, the framework
  presented here provides a simple approach --complementing existing
  theories, but specifically devised to account for the specificities
  of the recent empirical findings for plant communities-- to explain
  the collective emergence of diversification at a community level,
  and paves the way to further scrutinize the intimate entanglement of
  ecological and evolutionary processes, especially in species-rich
  communities.
\end{abstract}

\section{Introduction}
Community ecology studies how the relationships among species and
their environments affect biological diversity and its distribution,
usually neglecting phenotypic, genetic and evolutionary changes
\citep{Macarthur1984,Rosenzweig1995,Tilman-book}.  In contrast,
evolutionary biology focuses on genetic shifts, variation,
differentiation, and selection, but --even if ecological interactions
are well-recognized to profoundly affect evolution \citep{Ford1975}--
community processes are often neglected. Despite this apparent
  dichotomy, laboratory analyses of microbial communities and microcosms
  \citep{micro,Rapid2003,microcosm,friesen2004experimental,tyerman2005unparallel,herron2013parallel,Maclean,elena2003evolution,Cordero,spencer2007adaptive}
  as well as long-term field experiments with plant communities
  \citep{Tilman2006,Strauss2008} and vertebrates
  \citep{Rapid-lizards,Rapid-finches} provide evidence that species can rapidly (co)evolve and that
  eco- and evolutionary processes can be deeply intertwined even over
  relatively short (i.e. observable by individual researchers)
  timescales \citep{Ellner2011}.

Over the last two decades or so, the need to consider feedbacks between ecological
and evolutionary processes has led many authors to develop a framework to merge together the two fields
  \citep{Slatkin1980,Levin1992,Dieckmann1995,dieckmann1996dynamical,Law1997,Thompson,Dieckmann1999,doebeli2000evolutionary,doebeli2003speciation,Schluter,Loreau2004,Carroll,newest1,deangelis2005individual,Champagnat2006,newest2,Gravel2006,Aguiar2009,newest3,Hanski2012,doebeli2011adaptive,Tilman}.
  In particular, the development of quantitative trait models
  \citep{Fussmann2007} and the theories of adaptive dynamics
\citep{Geritz1997,Mesze1998} and adaptive diversification
\citep{Dieckmann1995,dieckmann1996dynamical,Law1997,Dieckmann1999,doebeli2000evolutionary,doebeli2003speciation,Champagnat2006},
  reviewed in \citep{Fussmann2007,doebeli2011adaptive}, has largely
  contributed to the rationalization of eco-evolutionary dynamics,
  shedding light onto non-trivial phenomena such as sympatric
  speciation and evolutionary branching \citep{doebeli2011adaptive}.

On the empirical side, the recent work by Zuppinger-Dingley
{\it et al.} on long-term field experiments of vegetation dynamics
appears to confirm many of the theoretical and observational
predictions \citep{Zuppinger}. This study provided strong evidence for the
emergence of \emph{rapid collective evolutionary changes}, resulting
from the selection for complementary character displacement and niche
diversification, reducing the overall level of competition and
significantly increasing the ecosystem productivity within a
relatively short time.  This result is not only important
for understanding rapid collective evolution, but also for designing
more efficient agricultural and preservation strategies. More
specifically, in the experimental setup of Zuppinger-Dingley and
colleagues, $12$ plant species of different functional groups were
grown for $8$ years under field conditions either as monocultures
or as part of biodiverse communities. Collecting plants (seedlings 
and cutlings) from these fields, propagating them in the laboratory, 
and assembling their offspring in new communities, it was possible 
to quantify the differences between laboratory mixtures consisting of 
plants with a history of isolation (i.e. from monocultures) and plants 
from biodiverse fields. While the former maintained essentially their 
original phenotypes, the latter turned out to experience significant 
complementary trait shifts --e.g. in plant height, leaf thickness, etc.-- 
which are strongly suggestive of a selection for phenotypic and niche 
differentiation \citep{Tilman} (see Fig. 1 therein). Furthermore, 
there were strong \emph{net biodiversity effects} \citep{Loreau2001}, 
meaning that the relative increase in total biomass production in laboratory
mixtures with respect to laboratory monocultures was greater for
plants from biodiverse plots than for plants coming from monocultures.
These empirical results underscore the need for simple
theoretical methodologies, in the spirit of the above-mentioned
synthetic approaches \citep{Slatkin1980,Tapper1992,dieckmann1996dynamical,doebeli2000evolutionary,doebeli2003speciation,Fussmann2007,Gravel2006,doebeli2011adaptive,Frey2011}. 
These approaches should explain the community and evolutionary
dynamics of complex and structured communities such as the ones analyzed in
\citep{Zuppinger}.

  The phenotypic differentiation observed in the experiments of
  Zuppinger-Dingley et al. might be partially
  rationalized within the framework of relatively simple
  deterministic approaches to eco-evolution such as adaptive dynamics
  (see e.g.
  \citep{Slatkin1980,Tapper1992,dieckmann1996dynamical,doebeli2000evolutionary,doebeli2003speciation,Fussmann2007,Gravel2006,doebeli2011adaptive}).
  In this context, diversification is the natural outcome of an
  adaptive/evolutionary process that increases fitness by decreasing 
  competition through trait divergence.
  
  However, it is not obvious what would be the combined effects in
  this simplistic version of adaptive dynamics of introducing elements
  such as sexual reproduction, space, and multi-species interactions
  that could play an important role in shaping empirical
  observations. Moreover, questions such as whether phenotypic
  differentiation occurs both above and below the species level (i.e.,
  within species or just between them), the possibility of long term
  coexistence of phenotypically equivalent species in the presence of
  strong competition (i.e., emergent neutrality), or the expected
  number of generations needed to observe significant evolutionary
  change remain unanswered and require a more detailed and specific
  modeling approach, within the framework of adaptive
  dynamics.

  Thus, our aim here is to contribute to the understanding of
  eco-evolutionary dynamics, emphasizing collective co-evolutionary
  aspects rather than focusing on individual species or pairs of them.
  For this purpose, we developed a simple computational framework
  --similar to existing approaches (see Discussion)--
  specifically devised at understanding the emerging phenomenology of
  the experiments of Zuppinger-Dingley {\it et al.} In particular, we
  propose an individual-based model, with spatial structure,
  stochasticity, sexual reproduction, mutation, multidimensional
  trait-dependent competition and, importantly, more than-two-species
  communities (in particular, possibly owing to analytical
  difficulties, relatively limited work has been published about more
  than three-species communities, which is crucial to achieve a
  realistic integration of ecological and evolutionary dynamics for
  natural communities; see however
  \citep{scheffer2006self,bonsall2004life,jansen1999evolving}).
  Furthermore, our method is flexible enough as to be easily
  generalizable to other specific situations beyond plant communities
  and can rationalize the circumstances under which phenotypic
  diversification and niche specialization may emerge using simple,
  straightforward rules.

\section{Results}

\subsection{Model essentials}

We construct a simple model which relies on both \emph{niche} based
approaches \citep{MacArthur-Levins1967,Chesson2000,Chase-Leibold2003}
and \emph{neutral} theories
\citep{Hubbell-Book,Age10,Volkov2003,Review-Padova}.  The former prioritize
 trait differences and asymmetric competition,
underscoring that coexisting species must differ in their
eco-evolutionary trade-offs, i.e., in the way they exploit diverse
limiting resources, respond to environmental changes,
etc., with each trade-off or ``niche'' choice implying superiority
under some conditions and inferiority under others
\citep{Macarthur1984,Tilman-book,Chesson2000,Chase-Leibold2003}.
Conversely, neutral theory ignores such asymmetric interactions by
making the radical assumption of species equivalence, and focuses on
the effects of demographic processes such as birth, death and migration.

Here, we adopt the view shared by various authors
\citep{Tilman2004,Reconciling,Loreau2011,Gravel2006} that niche-based
and neutral theories are complementary extreme views. In what follows, we
present a simple model that requires of both neutral and niche-based elements.
In particular, our model incorporates trade-off-based features such as
the existence of heritable phenotypic traits that characterize each
single individual. However, the impact of these traits on individual fitness 
is controlled by a model parameter, that can be tuned to make the process
	more or less dependent on competition, in the limit even mimicking neutral  
	(or ``symmetric'') theories \citep{Hubbell-Book,Age10}.

\begin{figure}[h!]
   \includegraphics[width=\textwidth]{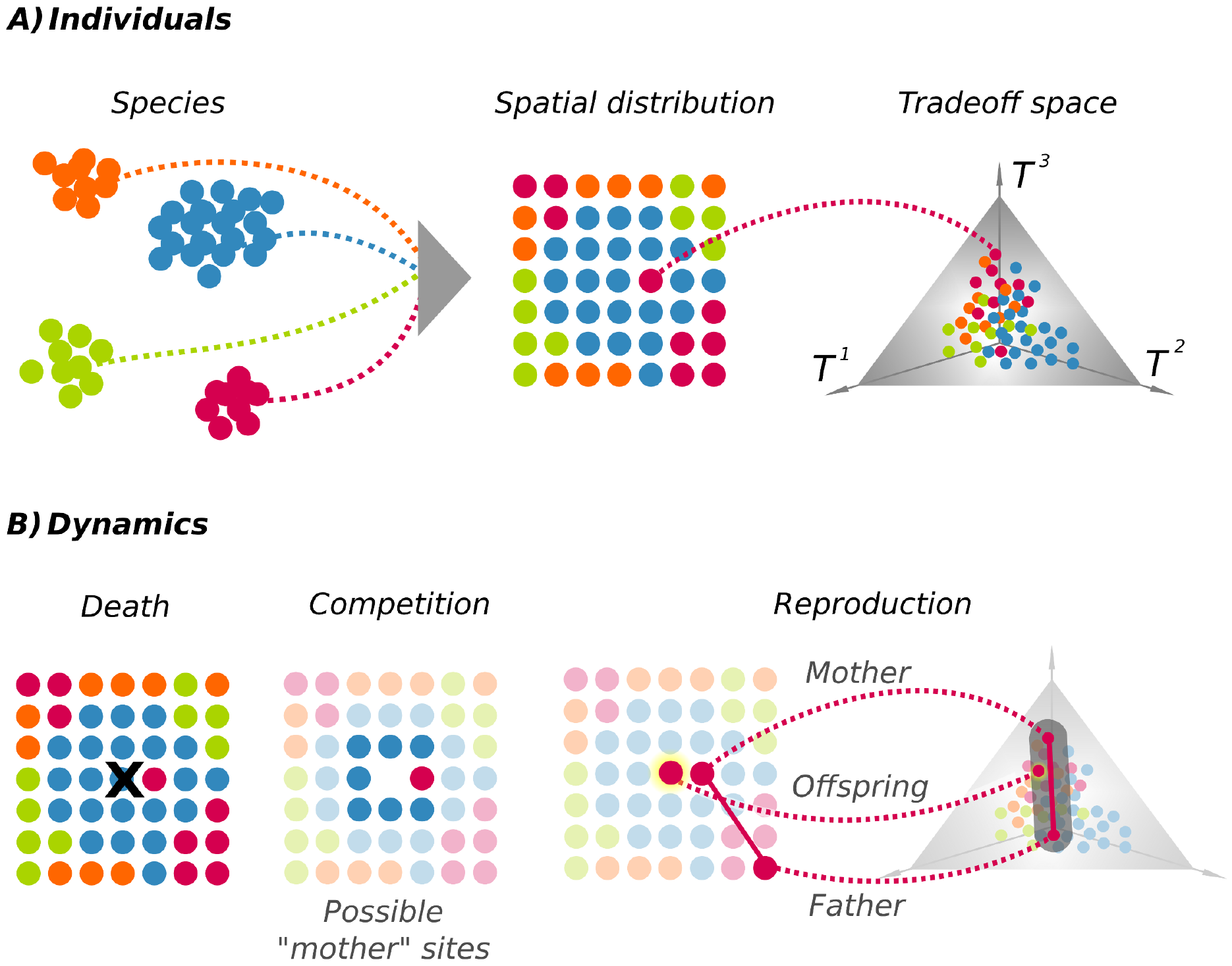}
  \caption{{\bf (Color online) Sketch of the model.}  {\bf(A)}
    Individuals of different species (different colors) compete for
    available resources in a physical space (two-dimensional square
    lattice), which is assumed to be saturated at all times. Each
    individual is equipped with a set of phenotypic traits that
    corresponds to a single point in the trade-off space. This is
    represented here (as a specific example) as an equilateral
    triangle (a ``simplex'' in mathematical terms) corresponding to
    the case of $3$ coordinates which add up to $1$ (e.g., fraction of
    the total biomass devoted to roots, leaves/stems and flowers,
    respectively \citep{Tilman}). For instance, a point close to vertex
    $T^1$ exploits better the limiting resource $1$ (e.g. soil
    nutrients) than another one near vertex $T^2$, but is less
    efficient at exploiting resource $2$ (e.g. light) than this latter
    one (see Methods).  {\bf (B)} Individuals die after one timestep,
    giving rise to empty sites; each of these is occupied by an
    offspring from a ``mother'' within its local
    neighborhood (consisting of $8$ sites in the sketch for clarity,
    although we considered also a second shell of neighbors in the
    simulations, i.e. a kernel of 24 sites).  The mother is randomly
    selected from the plants occupying this neighborhood in the
    previous generation, with a probability that decreases with the
    level of similarity/competition with its neighbors (see Methods).
    The implanted seed is assumed to have been fertilized by a
    conspecific ``father'' from any arbitrary random location,
    selected also with a competition-level dependent probability. The
    offspring inherits its phenotype from both parents; its traits can
    lie at any point (in the shaded region of the figure) nearby the
    the parental ones, allowing for some variation. For a given
      number of initial species $S$, two key parameters control the
      final outcome of the dynamics: $\beta$, characterizing the
      overall level of competition, and $\mu$, representing the
      variability of inherited traits. We fix most of the parameters
      in the model (lattice site, individuals within the
      competition/reproduction kernel, etc.) and study the dependence
      on $S$, $\beta$ and $\mu$.  }
\label{fig:1}
\end{figure}

The traits of each single individual are determined by quantitative
phenotypic values that can be regarded as the investment in specific
functional organs. For instance, the traits could represent the
proportion of biomass devoted to exploit soil nutrients (roots), light
(leaves and stems), and to attract pollinators and capture pollen
(flowers; see Fig. \ref{fig:1}). We then assume a hard limit --constant 
across generations-- to the amount of resources that
can be devoted to generate the phenotype, i.e. it is impossible to
increase all phenotypic values simultaneously. Thus each individual is
constrained to make specific trade-offs in the way it exploits
resources. Because similar values in the trade-off space entail
comparable exploitation of the same resource (e.g., water, light or
pollinators) similar individuals experience higher levels of
competition, which translates into a lower fitness. This can be
regarded as a frequency dependent selection mechanism providing an
adaptive advantage to exceptional individuals, able to exploit
available resources. Therefore, the ecological processes of
competition, reproduction, and selection lead to evolutionary shifts
in the distribution of phenotypic traits which feed back into
community processes, giving rise to integrated eco-evolutionary
dynamics.

\subsection{Model construction}
The basic components of the model are as follows (further details are
deferred to the Methods section).  We consider a community of
  individuals of $S$ different species, that are determined initially
  by mating barriers (i.e. a species is defined as a set of
  individuals that can produce fertile offspring
  \citep{Coyne2004}). Each individual occupies a position in physical
  space (represented as a saturated square lattice) and is
  characterized by the label of the species to which it belongs and a
  set of intrinsic parameters (i.e. trait values), specifying its
  coordinates in the ``trade-off space'' as sketched in
  Fig. \ref{fig:1} (see also \citep{rulands2014specialization,Tilman}.
All positions within the trade-off space are assumed to be equally
favorable \emph{a priori}. In what follows, we make a perfect
identification between the trade-offs of a given individual and its
phenotypic traits, which also determine the ``niche'' occupied by each
individual.  In principle, each individual, regardless of its species,
can occupy any positition in the trade-off space.  Positions near
  the center of the trade-off space (Fig. \ref{fig:1}) correspond to
  phenotypes with similar use of the different resources (i.e.,
  ``generalists''), while individuals near the corners specialize in
  the exploitation of a given resource (``specialists'').

Individuals are subjected to the processes of birth, competition for
resources, reproduction, descent with modification, and death.
Individuals are assumed to undergo sexual reproduction, as in the
experiments of \citep{Zuppinger} (implementations with asexual
reproduction are discussed later); they are considered to be
semelparous, so that after one simulation time step (i.e, a
reproductive cycle) they all die and are replaced by a new
generation. Importantly, demographic processes are strongly
dependent on phenotypic values. In particular, the main niche-based
hypothesis is that individual organisms with a better ``performance''
are more likely to reproduce than poorly performing ones.  To quantify
the notion of ``performance'', we rely on classical concepts such as
limiting similarity, competitive exclusion principle and
niche overlap hypothesis limiting similarity, competitive exclusion principle and
niche overlap hypothesis \citep{May1972,pianka1974}, which posit
that in order to avoid competition, similar species must differ in
their phenotypes. More specifically, our model assumes that the
performance of a given individual increases with its trait
``complementarity'' to its spatial neighbors \citep{pianka1974}, as
quantified by its averaged distance to them in trade-off space (see
Methods); i.e. the larger the phenotypic similarity among neighbors,
the stronger the competition, and the worse their performance.
Although the performance of a given individual depends on its
complementarity with its neighbors, the model is symmetric among
species and phenotypes; performance is blind to species labels and
does not depend on the specific location in the trade-off space.

The reproduction probability or performance of any given individual is
mediated by a parameter $\beta$ which characterizes the global level
of competitive stress in the environment (see Methods). In the limit
of no competition, $\beta=0$, the dynamics become blind to phenotypic
values and can be regarded as fully neutral, while in the opposite
limit of extremely competitive environments,
$\beta \rightarrow \infty$, niche effects are maximal and a relatively
small enhancement of trait complementarity induces a huge competitive
advantage.  Finally, a mother selected as described in the competition
process is assumed to be fertilized by a conspecific ``father'' in
the population (interspecies hybridization is not considered here)
which is also selected with the same reproduction probability function
based on its performance. The offspring inherits its traits from both
parents, with admixture and some degree of variation $\mu$ (see
Fig. \ref{fig:1} and Methods). This process is iterated for all
lattice sites and for an arbitrarily large number of reproductive
cycles, resulting in a redistribution of species both in physical and
in trade-off space. Species can possibly go extinct as a consequence of
the dynamics. In this version of the model, speciation is not
  considered, though it could be easily implemented by establishing
  a dependence of mating on phenotypic similarity, making 
  reproduction between sufficiently different individuals impossible \citep{Aguiar2009}.

\subsection{Computational results} 
\begin{figure}[h!]
\includegraphics[width=\textwidth]{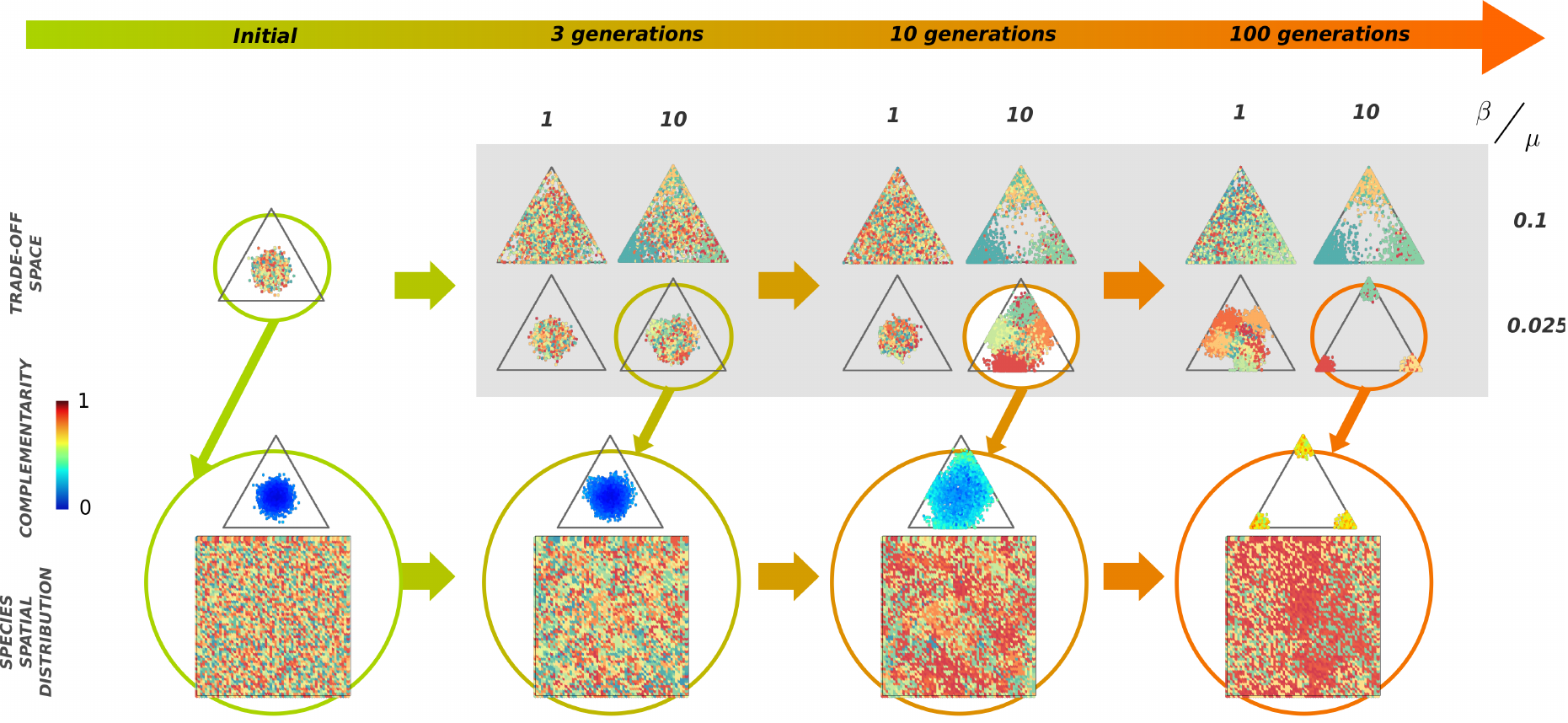}
\caption{{\bf (Color online) Illustration of the emergence of rapid
  phenotypic diversification}
  for a computational system of size
  $64 \times 64$ and $16$ species (labeled with different
  colors). {\bf(Top).} {\bf Phenotypic diagrams} measured at different
  evolution stages ($1$, $3$, $10$ and $100$ generations,
  respectively) for different values of the two parameters: level of
  competition $\beta$ ($1$ for the case of low competition and $10$
  for strong competition) and variation in inherited traits $\mu$
  ($0.1$ for large variation and $0.025$ for small variation). In all
  cases, phenotypic differentiation among species is evident even
  after only $10$ generations. In the long term ($100$ generations)
  species diversification and specialization is most evident for small
  $\mu$ and large $\beta$; in this last case, different species
  (colors) can coexist for large times in the same region/corner of
  trade-off space.  {\bf (Central).} {\bf Complementarity diagrams}
  representing the values of averaged local complementarity for all
  individuals of any species for small $\mu$ ($0.025$) and large
  $\beta$ ($10$). Individuals with small complementarity (i.e. under
  strong competition with neighbors) disappear in the evolutionary
  process, while communities with high degrees of local
  complementarity are rapidly selected. {\bf (Bottom).} {\bf Spatial
  distribution of species} for different number of generations.  As
  a result of the eco-evolutionary dynamics, anti-correlated patterns
  --in which neighboring plants tend to be different-- emerge (note
  that colors represent species assignment and do not reflect
  phenotypic values)}
\label{fig:2}
\end{figure}

\begin{figure}[h!]
\includegraphics[width=0.9\textwidth]{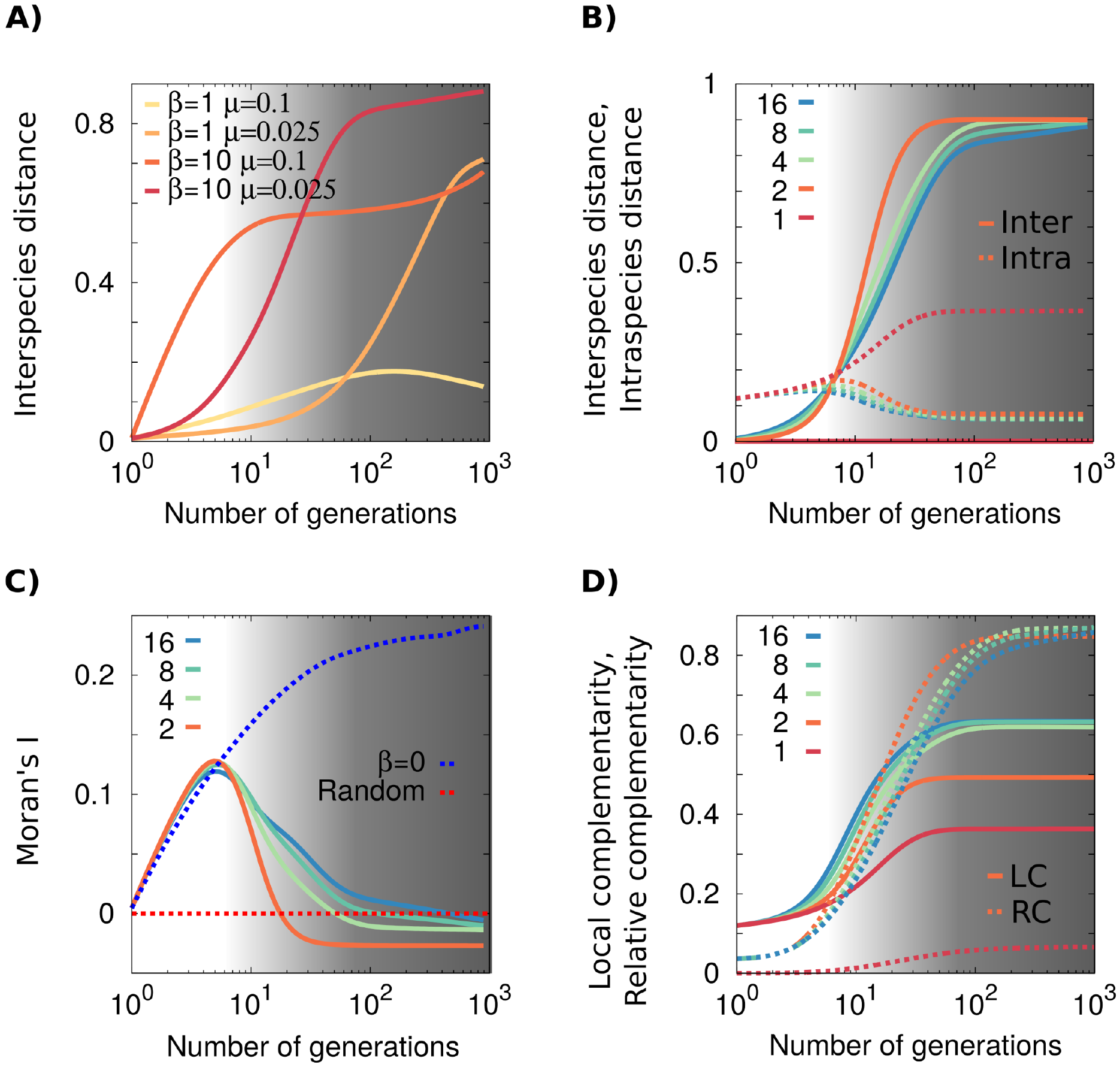}
  \caption{{\bf (Color online) Measurements of different biodiversity
      indices.}  {\bf(A)} {\bf{Phenotypic distances among species}}
    grow systematically during the eco-evolutionary
    process, reflecting a clear tendency towards species
    differentiation (same sets of parameter values as in
    Fig. \ref{fig:2}, $S=16$). Differentiation is faster for
    relatively small values of trait variability $\mu$ and large
    values of the competitive stress $\beta$. {\bf(B)}
    {\bf{Phenotypic differentiation among and within species}}. While
    interspecies distances grow in time for all values of $S$
    and converge to similar values on the long term, intraspecific phenotypic 
    variability is much larger on the long term
    for monocultures than for biodiverse mixtures. {\bf(C)}
    {\bf{Phenotypic similarity among close neighbors.}} Moran's index
    ($I$) for $\beta=10$ and different values of
    $S$ as well as for $\beta=0$ and for a random distribution
    (i.e. in the absence of spatial interactions). The value of $I$
     tends to $0$ for random distributions, is positive for
    $\beta=0$, and tends to small negative values for $\beta\ne 0$.
    Whenever competition depends on the phenotypic values
    (i.e., $\beta>0$) the system avoids close cohabitation of individuals
    of the same species. This negative spatial autocorrelation results in $I<0$; 
    in all cases, $\mu=0.025$. {\bf(D)} {\bf{Averaged
        local and relative complementarity}} in the community increase
    with time and reach larger values for more biodiverse communities.
   The phenotypic differentiation among individuals is greater both among
   close neighbors and at the global scale as the number os species $S$ increases. 
   In all plots, parameters are $L=64$ and, unless it is specified,
    $\beta=10$ and $\mu=0.025$; curves are averaged over at least
    $10^3$ runs; shaded light grey areas stand for times during which
    extinction tends to occur causing $S$ to decrease (see S10 Appendix in Supporting Information for details), 
    while in dark grey ones the system tend to stabilize at a given final number of species.}
\label{fig:3}
\end{figure}

Simulations are started with individuals of $S$ different species
(e.g. $S=16$) randomly distributed in space. In the initial
  conditions, the traits of all individuals are a sample from a common
  Gaussian distribution centered around the center of the simplex (note that
  as shown in the S5 Appendix in Supporting Information, results do not depend on the particular choice 
  of initial conditions). Statistical patterns emerging from the eco-evolutionary 
dynamics described above are analyzed as a function of the number of generations 
and as a function of the number of species $S$, for different values of the two
free parameters: the overall level of competition $\beta$ and the
variability of inherited traits $\mu$. Results are illustrated in
Fig. \ref{fig:2} showing \emph{(i)} phenotypic diagrams (top row)
specifying the position of each single individual and its species in
the trade-off space for different parameter values and evolutionary times
\emph{(ii)}; values of complementarity for all individuals (central
row) in the trade-off space, and \emph{(iii)} the spatial distribution
of individuals and species (bottom row).  Finally, several
biodiversity indices are reported in Fig. \ref{fig:3}.

\subsubsection{Species differentiation} As illustrated in
Fig. \ref{fig:2} (shaded area), different distributions of individuals
in the trade-off space appear depending on the specific values of
$\beta$ and $\mu$. Visual inspection reveals the emergence of rapid
phenotypic differentiation, i.e. segregation of colors in trade-off
space after a few (e.g. $10$) reproductive cycles.  The segregation is
much more pronounced for relatively small variability
(e.g. $\mu=0.025$) and large competitive stress (e.g. $\beta=10$).
This is quantified (see Fig. \ref{fig:3}A) by the average interspecies
distance (see Methods), whose specific shape depends on parameter
values.  As shown in Fig. \ref{fig:3}B, the fastest growth is obtained
for $S=2$, but the curves converge to a constant value (mostly
independent of $S$) after a sufficiently large number of generations.
Moreover, as shown in the central row of Fig. \ref{fig:2} the
complementarity --averaged over all individuals in the community (see
Methods)-- also grows during the course of evolution (i.e. colors
shift from blue to yellowish).  Observe in Fig. \ref{fig:2}. that, for
asymptotically large evolutionary times, there is a tendency for all
species to cluster around the corners of the trade-off space,
suggesting that the optimal solution to the problem of minimizing the
competition with neighbors corresponds to communities with highly
specialized species. This specialization does not
occur in monocultures ($S=1$), as sexual mating pulls the species together
and avoids significant phenotypic segregation.

\subsubsection{Emergence of local anti-correlations} The high level
of phenotypic specialization observed after large evolutionary times
for large competition stress and small variability, might seem in
contradiction with the overall tendency to niche differentiation.  In
other words, most of the trade-off space becomes empty in this case,
while individuals aggregate at the (highly populated) corners.  The
answer to this apparent conundrum is that similarly specialized
individuals have a statistical tendency to avoid being spatial
neighbors.  Indeed, as qualitatively illustrated in the lowest right
panel of Fig. \ref{fig:2}, extreme specialization is accompanied by a
tendency to diminish spatial clustering, i.e. to create spatial
anti-correlations within each species.  This tendency --which stems
from intraspecific competition and opposes to the demographic
tendency of similar individual to cluster in space-- is quantitatively
reflected by negative values of Moran's index $I$
(see Fig. \ref{fig:3}C and Methods). Note also that $I$ and thus the 
spacial distribution of species, is radically different in the presence 
and in the absence of competition (i.e. for $\beta \neq 0$ and $\beta =0$, 
respectively) as can be seen in Fig. \ref{fig:3}C.  In the absence 
of competition, species are distributed randomly forming aggregated 
spatial clusters without competition-induced local anti-correlations.

\subsubsection{Intraspecific diversity} This quantity is defined as
the mean ``complementarity'' among all pairs of conspecific
individuals in the community, and illustrates the level of phenotypic
diversity within species. As shown in Fig. \ref{fig:3}B, the
intraspecific diversity is much larger for monocultures.  In
monocultures, neighbors are obviously conspecific and the only
available mechanism to reduce overall competition is to increase
intraspecific diversity. Therefore, as a general result, monocultures
tend to enhance their intraspecific phenotypic distances, while
biodiverse communities tend to enhance phenotypic
differentiation among species but result in more similar conspecifics.

\subsubsection{Local complementarity} Fig. \ref{fig:3}D shows the
evolution of the mean complementarity of individuals respect to its
spatial neighbors.  This averaged \emph{local} complementarity (LC)
controls the dynamics and the actual reduction in the level of
competition for a given spatial distribution, and is much larger for
mixtures than for monocultures (it grows monotonously with $S$ and
saturates at a maximal value).

\subsubsection{Global complementarity}
Similarly, we can measure ``\emph{global}'' complementarity (GC),
i.e. the average phenotypic distance among all individuals in the
experiment, regardless of their spatial coordinates, after a given
number of generations.  Additionally, we measured
$\mathrm{GC}_\mathrm{intra}$ (resp.  $\mathrm{GC}_\mathrm{inter}$)
which is $\mathrm{GC}$ averaged only over individuals of the same
(resp. different) species (see Methods). In Fig. \ref{fig:3}D we
present results for the \emph{relative} complementarity
$\mathrm{RC}=\mathrm{GC}_\mathrm{inter}-\mathrm{GC}_\mathrm{intra}$,
which is a measure of the averaged difference in the level of
competition between randomly sampled non-conspecific and conspecific
individuals, respectively. Observe that the RC is larger for mixtures
than for monocultures, $\mathrm{RC}(S>1) > \mathrm{RC}(S=1)$, and that
it grows faster in time for smaller values of $S$ (e.g. $S=2$), but
reaches almost equal constant values after a sufficiently large number
of generations.

\subsubsection{Emergent neutrality} As
  illustrated in Fig. \ref{fig:2}, different species with very similar
  trait values can coexist (e.g. yellow and orange species at the
  right corner of the phenotypic diagram for $\mu=0.025$ and
  $\beta=10$ in Fig. \ref{fig:2}) even after many generations.  Such a
  coexistence emerges spontaneously and although it is transitory it
  can last for arbitrarily long times provided that the system size is
  sufficiently large. From an ecological point of view, these species
  can be regarded as functionally equivalent
  as they occupy the same niche region (see S6 Appendix in the Supporting Information for
  a detailed analysis of the stability and coexistence time of
  such species).
 
\subsubsection{Model variants} To investigate the generality of our
findings, we also explored whether the main conclusions are robust
against some constraints of the implementation.  We briefly explain
the variants we took into account below. i) \emph{Non-symmetrical
  phenotypic trade-offs:} as a first step, we assumed that not all
positions in the trade-off space are equally rewarding \emph{a
  priori}: individuals in certain regions of the trade-off space have
larger reproduction probabilities than others.  Non-symmetrical
trade-offs lead to very similar results as above, confirming the
robustness of our conclusions (see S7 Appendix in Supporting Information). ii)
  \emph{Asexual reproduction}: as shown in S8 Appendix in Supporting Information, for
  communities of individuals able to reproduce asexually
  (i.e. assuming transmission of traits only from the mother) the
  outcome of the model is different: individuals tend to diversify,
  but such diversification occurs even within species
  (i.e. intra-specific diversification is much larger than in the
  sexual case); in other words, since there is no admixing of the
  phenotypic traits through reproduction, diversification occurs
  within maternal lineages, rather than at the species level.  iii)
\emph{Long-distance dispersal and competition}: we also studied the
case in which dispersal and competition are long-distance processes
and affect the whole community and not only close neighbors; as shown
in S3 Appendix in Supporting Information the phenomenology reported above remains quite
  similar, even if in this well-mixed case the co-existence of
  phenotypically-equivalent species is less likely (owing to the lack
  of spatial separation). iv) \emph{Effect of the competition
    kernel:} the specific form of the competition kernel can play a
  crucial role in the formation of species clusters in phenotypic
  space \citep{Geritz1999, Adler-Mosquera2000, Pigolotti2007,
    Ramos2008, Hernandez2009, Barabas2013, Lampert2014}; in the S9
  Appendix of the Supporting Information we explore different kernel functions and show that our
  results are robust against changes in the mathematical expression of
  competition.

\section{Discussion} 

In the present paper, we have developed a parsimonious modeling
approach to integrate important ecological and evolutionary
processes.  In particular, we focused on understanding rapid
phenotypic diversification observed in complex biological communities
of plants such as those recently reported by Zuppinger-Dingley
{\emph{et al.}} in long-term field
experiments \citep{Zuppinger,Tilman}. 

Our model blends standard community processes, such as reproduction,
competition or death, with evolutionary change (e.g., descent with
modification); i.e.  community and evolutionary dynamics are coupled
together, feeding back into each other.  Over the last decades,
attempts to integrate ecological and evolutionary dynamics have been
the goal of many studies (see
e.g. \citep{Dieckmann1995,dieckmann1996dynamical,Law1997,Thompson,Dieckmann1999,
  doebeli2000evolutionary,doebeli2003speciation,Schluter,Loreau2004,Carroll,Strauss2008,newest1,
  deangelis2005individual,Champagnat2006,Aguiar2009,Hanski2012,doebeli2011adaptive,Frey2011,rulands2014specialization}).
In particular, a basic algorithm for modeling eco-evolutionary
dynamics as a stochastic process of birth with mutation, interaction,
and death was proposed in \citep{Dieckmann1995} and much work has been
developed afterwards to incorporate elements such as spatial effects
and different types of interspecies interactions
\citep{doebeli2003speciation}.

Rather than providing a radically different framework, our model
constitutes a blend of other modeling approaches in the literature of
eco-evolutionary processes, and in fact it shares many ingredients
with other precedent works, specially with the theory of adaptive
dynamics \citep{doebeli2011adaptive,Fussmann2007}.  For instance,
Gravel et al. \citep{Gravel2006} also considered a spatially-explicit
individual-based model with trait-dependent competition. 
However, our work has been specifically devised to shed light on the
  experimental findings of Zuppinger et al.\citep{Zuppinger}, and puts the emphasis on communities with
arbitrarily large number of species, while usually the focus is on the (co-)evolution of pairs of
species (e.g. predator-prey, host-parasite, etc.) or speciation/radiation of individual species.
Finally, our modelling approach is sufficiently general as to be flexible to be adapted to other 
situations with slightly different ingredients. We explored some of these possible extensions in
some Appendices (S3,S4,S7,S8,S9) in Supporting Information (e.g long-distance dispersal, asexual reproduction,
etc.), but other studies can be built upon the work laid here in a relatively simple way.

The present model relies on a number of specific assumptions, two of
which are essential in that they couple community and evolutionary
dynamics: i.e.  (i) demographic processes are controlled by
competition for resources which is mediated by phenotypic traits and
(ii) successful individuals are more likely to transmit their
phenotypes to the next generation with some degree of variation.
These two ingredients are critical for the emerging phenomenology.
For instance, in the absence of competition (i.e. $\beta=0$)
reproduction probabilities are identical for all individuals, implying
that the model becomes neutral, and the evolutionary force leading to
species differentiation vanishes (see S4 Appendices in Supporting Information).  On the other
hand, variation in inherited traits is necessary to allow for the
emergence of slightly different new phenotypes and the emergence of
drifts in trade-off-space. Although these constraints
might be regarded as limiting, we deem them biologically
realistic and do not think they hamper the predictive power of
our model.  Most of the remaining ingredients, such as the existence
of a saturated landscape, semelparity (i.e. non-overlapping
generations), the specific form in which we implemented initial
conditions, competition, dispersion, selection, inheritance linked to
phenotypic characters rather than to a genotypic codification,
etc. can be modified without substantially affecting the results. This
flexibility could make the description of other type of communities
possible with minimal model variations.  Similarly, the model could be
extended to incorporate phenotype-dependent reproductive barriers 
(and thus speciation) and the possibility of interspecies hybridization by making
reproduction a function of phenotypic distance and relaxing its
dependency on species labels.

  In addition to rapid phenotypic diversification, the experiments
  of Zuppinger et al. found an enhancement of the overall productivity
  in mixtures of diverse plants with respect to monocultures of the
  same plants \citep{Zuppinger}.  Our model cannot be used to directly
  quantify such ``biodiversity effects'' \citep{Loreau2001}, as we
  assume a fully saturated landscape and there is no variable that
  accounts for total biomass production.  However, in principle, under
  the hypothesis that larger trait complementarities correlate with
  greater resource capture and biomass production, the observed
  increase of relative complementarity in mixtures (see Fig. 3) could
  be used as a proxy for biodiversity effects. Observe, nonetheless,
  that the previous assumption might by wrong (or incomplete) as productivity
  can be profoundly affected by other factors such as, for instance,
  positive interactions between similar species, not modeled here, and
  more sophisticated approaches --see \citep{hooper2005effects,
    duffy2007functional,cardinale2009towards, reiss2009emerging,
    grace2016integrative}-- are necessary to validate this hypothesis.
  In the future we plan to modify our model to represent non-saturated
  landscapes and more detailed ecological dynamics, allowing for
  explicit analyses of biodiversity-productivity relationships.

Beyond explaining most of the empirical observations in
\citep{Zuppinger}, our model leads to some far-reaching predictions
(some of them already shared by existing theories);
one of the most remarkable ones is that optimal exploitation of
resources comes about when the full community evolves into a reduced
number of highly specialized species --the exact number depending
on the dimensions of the trade-off space-- that coexist in highly
dispersed and intermixed populations. Such specialization might 
be unrealistic in the case in which all traits in trade-off space are essential 
for survival, and thus the convergence toward perfect specialization is capped.  
In any case, this result is congruent with the niche dimension hypothesis
\citep{Hutchinson}, that postulates that a greater diversity of niches
entails a greater diversity of species, i.e. a larger number of
limiting factors (and thus of possible trade-offs) leads to richer
communities \citep{harpole2007}. However, this outcome might be
affected by perturbations (migration, environmental variability, etc)
which could be easily implemented in our model, and could prevent real
communities from reaching the asymptotic steady state predicted
here. It is also noteworthy that the resulting highly specialized
species can be phenotypically equivalent, and a set of
them can occupy almost identical locations in the trade-off space.
Such species equivalence appears spontaneously, and supports the views
expressed by other authors that ``emergent neutrality'' is a property
of many ecosystems \citep{Allan_neutrality,vanNes,Simone}. In future
work we will explore the possibility of phase transitions separating
an ecological regime based on the coexistence of multiple highly
specialized species from an ecosystem dominated by generalists and the
conditions under which each regime emerges.

  Beyond phenotype-dependent mating, upcoming studies will extend
  our approach to address communities where collective diversification
  phenomena based on both competition and cooperation are known to
  emerge (see e.g. \citep{Cordero}), as well as investigate the
  evolution of communities with distinct types of interacting species
  such as plant-pollinator mutualistic networks. This research will
hopefully complement the existing literature and help highlighting the
universal and entangled nature of eco-evolutionary processes.

\section{Methods}
\subsection{Model implementation}
We implemented computer simulations in which each individual plant,
$i$, is fully characterized by (see also Fig. \ref{fig:1}): \emph{(i)}
a label identifying its species, \emph{(ii)} its coordinates in the
physical space, and \emph{(iii)} a set of real numbers specifying its
phenotypic traits. In these simulations, time can be implemented
either as discrete/synchronous updating or continuous/sequential
updating without significantly altering the results. \emph{Species--},
we consider a fixed number of species, labeled from $1$ to $S$; while
the emergence of new species is not considered here, some of them may
become extinct along the course of evolution.  \emph{Physical space--}
We consider a two-dimensional homogeneous physical space described by
a $L\times L$ square lattice, assumed to be saturated at all times, in
which the neighborhood of each individuals is determined by the
closest $K$ sites (in our simulations, we took $L=64$ and $K=24$).
\emph {Phenotypic traits and trade-off space--} As energy and resources
are limited, each individual plant needs to make specific
choices/trade-offs on how to allocate different functions.  The way we
implement the ``trade-off space'' is inspired in the field of
multi-constraint (non-parametric) optimization that it is called
Pareto optimal front/surface \citep{Pareto}; it includes the set of
possible solutions such that none of the functions can be improved
without degrading some other.  Thus, the phenotype of any individual
can be represented as a trade-off equilibrium, a point in this space
and encapsulated in a set of real numbers
$\mathbf{T}=(T^1,T^2,...,T^{n})$ (all of them in the interval
$[0,1]$), such that $\sum_{k=1}^{n} T^k=1$ where $n$ is the number of
trade-offs (see Fig. \ref{fig:1} and \citep{Tilman}). All positions
within the trade-off space are equivalent a priori, although this
requirement can be relaxed.  \emph{Competition for resources--} The
trait ``complementarity'' between two individuals $i$ and $j$ is
quantified as their distance in the trade-off space: $c_{ij}
=\sum_{k=1}^n |T^k(i)-T^k(j)|/n$, which does not depend on species
labels. The averaged complementarity, (or simply ``complementarity'')
over all the neighbors $j$ of individual $i$ is $C_i = \sum_{j \in
  n.n.(i)} c_{ij}/K$.  \emph{Complementarity-based dynamics--} Each
timestep, every individual is removed from the population; the
resulting vacant site $i$ is replaced by an offspring of a potential
mother plant $j$ which is selected from the list of $K$ local
neighbors of the vacant site with a given probability
$P_{\mathrm{mother}}(j)$. This probability controls the dynamical
process; we assume it to increase as the mother's trait
complementarity $C_j$ increases (i.e. as its effective competitive
stress diminishes): $ P_{\mathrm{mother}}(j)= {e^{\beta
    C_j}}/{\sum_{j' \in n.n.(i)} e^{\beta C_{j'}}}~$, where the sum
runs over the set of $K$ neighbors of $i$; $e^{\beta C_j}$ is the
``performance'' of individual $j$ and $\beta$ is a tunable
``competition parameter'' controlling the overall level of competitive
stress in the community.  Once the mother has been selected, the
father is randomly chosen from all its conspecific individuals $l$ in
the community, with a probability proportional to their performance,
${e^{\beta C_l}}$. In other words, individuals with lower competition
pressure are more likely to sire descendants both as females and as
males.  \emph{Inheritance, admixture and variation of phenotypes--}
The traits of each single offspring are a stochastic interpolation of
those of both parents with the possibility of variation:
$T^k_\mathrm{new} = \eta T^k_\mathrm{mother} +
(1-\eta)T^k_\mathrm{father} + \xi^k$, for $k=1,...,n$, where $\eta$ is
a random variable (uniformly distributed in $[0,1]$) allowing for
different levels of admixture for each offspring, and $\xi_k$ are
(Gaussian) zero-mean random variables with standard deviation $\mu$,
  a key parameter that characterizes the variability of inherited
  traits. To preserve the overall constraints $T^k \in [0,1]$ and
  $\sum_k T^k=1$, mutations are generated as $\xi^k=(r^k-r^{k+1})$,
  where $\{r^1=r^{n+1},...,r^n\}$ are independent Gaussian random
  variables with zero-mean and standard deviation $\mu/\sqrt{2}$; in
  the rare case that $T^k_\mathrm{new} <0$ (resp. $>1$), we set it to
  $0$ (resp. to $1$) and added the truncated difference to another
  random trait.

\subsection{Biodiversity indices}
\emph{The centroid of species $s$} is
${\bf{B}}(s)=\{B^1(s),...,B^n(s)\}$, with $B^k(s)=\sum_{i} T^k(i)/n_s$
for each trait $k$, where $i$ runs over the $n_s$ individuals of
species $s$.  \emph{Interspecies distance}: is the distance between
the centroids of two different species $s$ and $s'$ in the trade-off
space $d_{s,s'}=\sum_{k}|B^k(s)-B^k(s')|/n$, averaged over all
surviving species.  {\emph{Intraspecific distance}} is the average
distance in trade-off space between all pairs of individuals of a
given species $s$, $d_s=\sum_{i,j \in s}c_{ij}/n_s(n_s-1)$ averaged over
all surviving species.  {\emph{Local complementarity}} is the the mean complementarity of individuals to their spatial neighbors,
$\mathrm{LC}=\sum_{i}(\sum_{j \in n.n.(i)}c_{ij}/K)/N$ where $N$ is
the total number of individuals and $K$ is the number of local
neighbors.  {\emph{Global complementarity}} is the complementarity averaged
over all pairs of individuals regardless of their relative positions in physical space, $\mathrm{GC}=\sum_{i,j\neq i}c_{ij}/(N(N-1))$. 
Similarly $\mathrm{GC}_\mathrm{inter}$ is the averaged complementarity between
individuals of different species  and $\mathrm{GC}_\mathrm{intra}$ is the
averaged complementarity between conspecific individuals. In the case of monocultures, $\mathrm{GC}_\mathrm{inter}(S=1)$
is measured from two different/independent realizations. 
{\emph{Relative complementarity}, $\mathrm{RC}=\mathrm{GC}_\mathrm{inter}-\mathrm{GC}_\mathrm{intra}$, is a
measure of the averaged difference in the level of competition
between randomly sampled conspecific and non-conspecific
individuals. 
\emph{Moran's index}: is a measure of spatial
correlations between neighbors; it is negative when neighbors tend
to belong to different species (see S2 Appendix in Supporting Information).

\section{Acknowledgements} 
 We acknowledge support from the Spanish MINECO project
  FIS2013-43201-P.  RRC is supported by the Talentia Program (Junta de
  Andalucía / EC-FP7 COFUND – Grant Agreement 267226).

\newpage
\section*{Supporting Information}

\subsection*{S1. Local and global trait complementarities}
\label{Local and global complementarities}

We distinguish three different measures regarding averaged
complementarities (see the sketches in Fig. \ref{fig:Appendix_BE}):
\emph{i) }{\bf Local complementarity (LC)}, defined as the mean
phenotypic distance between an individual and its spatial neighbors;
\emph{ii) } {\bf Intraspecific global complementarity
  (GC$_\mathrm{intra}$)}, defined as the mean phenotypic distance
between all pairs of individuals within the same species (and then
averaged over species); and, additionally, \emph{iii)} {\bf
  Interspecific global complementarity (GC$_\mathrm{inter}$)},
corresponding to the the mean phenotypic distance between all pairs of
individuals consisting of individuals of two different species
(averaged over all species).  While local measurements capture the
effect of spatial correlations, global ones are useful to characterize
the evolution of the whole community in trade-off space. In analogy
with the experimental setup of Zuppinger-Dingley \emph{et al.}
\citep{Zuppinger}, we performed computer simulations using both
monocultures and mixtures. For the case of monocultures,
GC$_\mathrm{inter}$ was estimated taking individuals coming from two
independent realizations of the simulations.  Zuppinger et
al. gathered seeds from surrounding populations and grew them together
in their experimental set-up.  This removed any cumulative,
trans-generational effect of spatial correlations.  In this way, GC
measurements (Fig. \ref{fig:Appendix_BE} B) constitute better proxies
(as compared to LC) to contrast our results with global
community (biodiversity) effects in \citep{Zuppinger}.

\begin{figure}[h!]
 \includegraphics[width=0.8\textwidth]{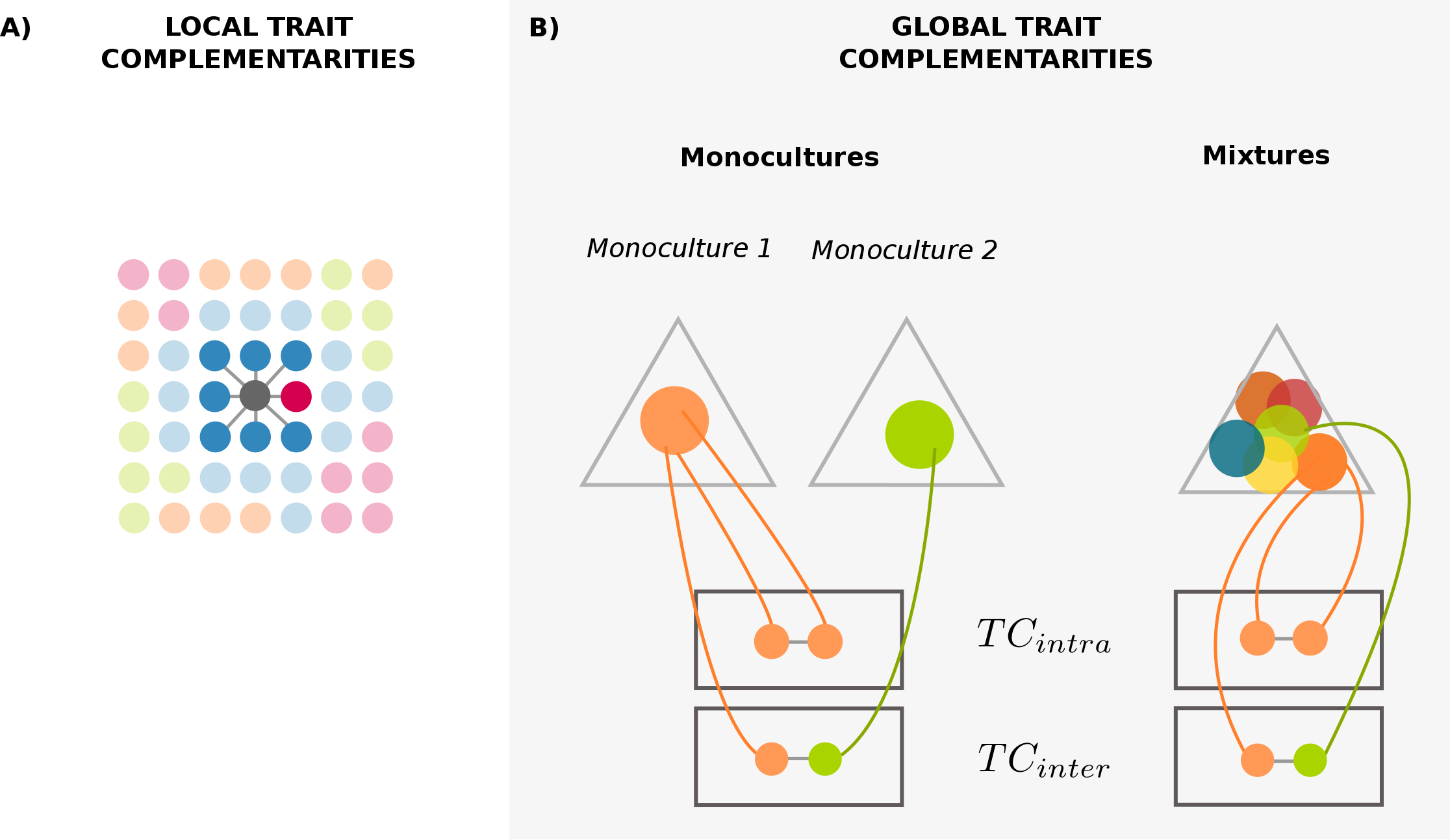}
 \caption{\textbf{(Color online) complementarity measurements:}} A)
 Local complementarity (i.e. mean phenotypic distance between
 spatially close neighbors) reveals the effect of spatial
 correlations; B) Intraspecific and interspecific global
 complementarities (i.e. mean phenotypic distance between individuals
 of the same and of different species, respectively, regardless of
 their spatial location).  In all cases, we performed computer
 simulations using monocultures and mixtures.
\label{fig:Appendix_BE}
\end{figure}

Complementing the results presented in Fig. 3 of the main text, Fig. \ref{fig:Appendix_LGTC}
shows measurements for LC and GC$_\mathrm{intra/inter}$ for different
initial number of species $S$ after $10$ and $1000$
generations. We observe that, even though complementarities become
almost independent of $S$ at $t=1000$ (due to the extinction of some
species), transient measurements at $t=10$ clearly show that
communities with fewer species exhibit higher values of GC
and, consequently, reach the stationary state faster. Biodiversity delays 
the process because several species simultaneously compete for empty niches.  
On the other hand, LC is inversely correlated with $S$, i.e., individuals tend to 
be more phenotypically similar to {\'a}s in less biodiverse communities.

\begin{figure}[h!]
\centering
 \includegraphics[width=0.8\textwidth]{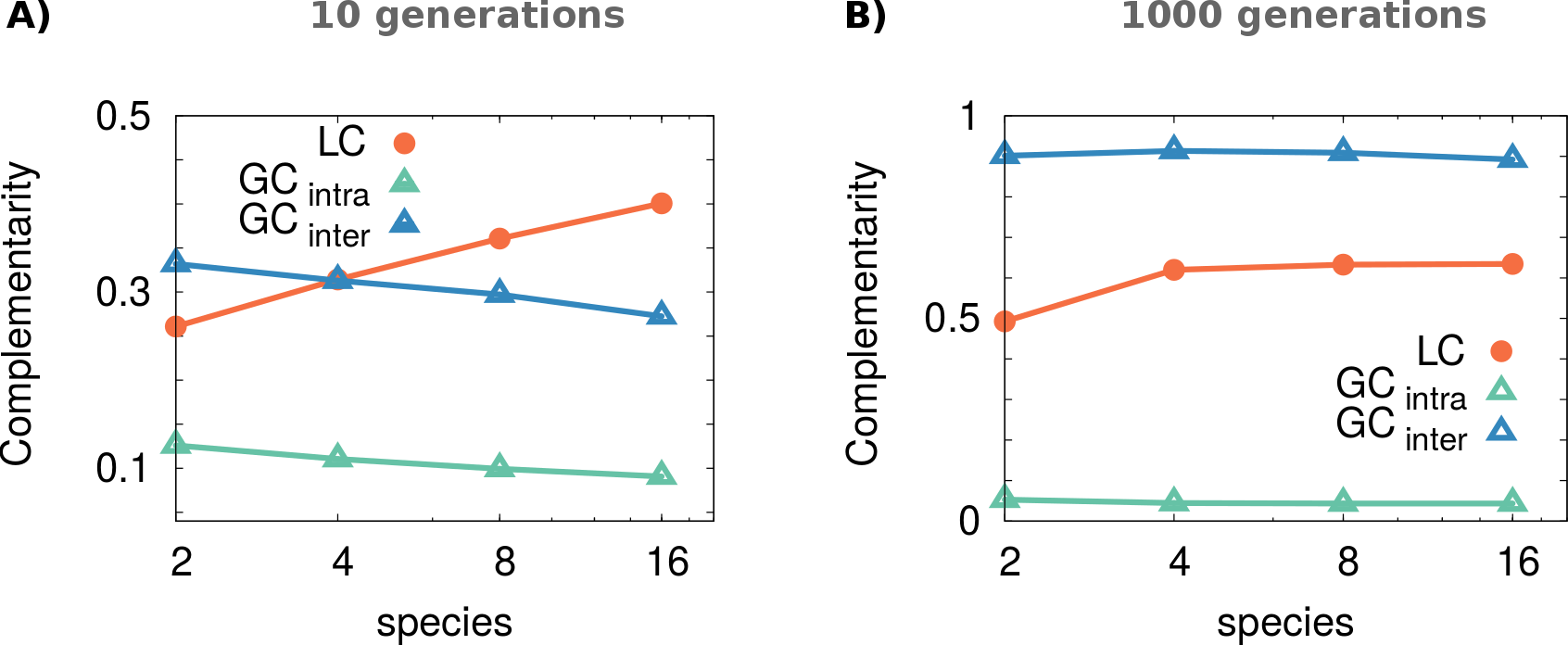}
 \caption{\textbf{(Color online) Local and global complementarities
     after A) $10$ and B) $1000$ reproductive cycles}, plotted as a
   function of the initial number of species $S$ in the
   community. Parameter values (see main text): system size $L=64$,
   competition $\beta=10$ and variability $\mu=0.025$.}
\label{fig:Appendix_LGTC}
\end{figure}

\subsection*{S2. Moran index}
\label{Moran index}
The Moran's index \citep{Moran1950} quantifies the likelihood of an
individual to be surrounded by individuals of the same species. When
Moran's index is negative, individuals are less likely to be close to
their co-specifics than what would be expected by pure chance, while
positive values indicate spatial clustering of
species. Mathematically, given a species $s$ we compute its Moran's
index $I_s$ as
\begin{equation}
 I_s=\frac{\sum_{i\in s} \sum_{j\in n.n.(i)} (X_s^i-\bar{X}_s)(X_s^j-\bar{X}_s)}{K\sum_{i\in s} (X_s^i-\bar{X}_s)^2},
\end{equation}
where $K$ is the number of local neighbors (kernel size), and $X_s^i$ is
a variable such that $X_s^i=1$ when the specie of $i$ is equal
to $s$ and $X_s^i=0$ if it is different, with $\bar X_s$ the density of individuals of
species $s$. Finally, we obtain the total index averaging over
species, $I=\sum_{i=1}^{S} I_s/S$. As a result, positive, zero, and
negative values of $I$ correspond to positive spatial correlation, random,
and anti-correlation of species, respectively.

\subsection*{S3. Long-distance dispersal and competition}
\label{mean-field}

We have also studied well mixed (or ``fully connected'' ) communities,
in which \emph{ i)} there is frequent long-distance dispersal (so that both
progenitors of the new offspring can be located at any site in space)
and, additionally, \emph{ii)} each individual competes with the rest
of the community, i.e. all individuals behave as nearest neighbors.

As illustrated in Fig. \ref{fig:Appendix_mean-field}, all the
previously reported phenomenology is still present in this ideal mean-field
scenario. As a matter of fact, phenotypic differentiation seems to occur faster than
when spatial distribution is conditioned by local dispersal (see Fig. 3). In other words, 
long-distance dispersal and global competition drive evolution faster than local
dynamics. This is a consequence of enhanced competition, which
increases the relative fitness of better performing individuals. Another important
difference is that, under mean field conditions, equivalent taxa cannot occupy
different spatial locations and are forced to compete with each other. 
Consequently, coexistence of species with similar traits is much less
likely than in spatially-explicit communities.

\begin{figure}[h!]
\centering
 \includegraphics[width=\textwidth]{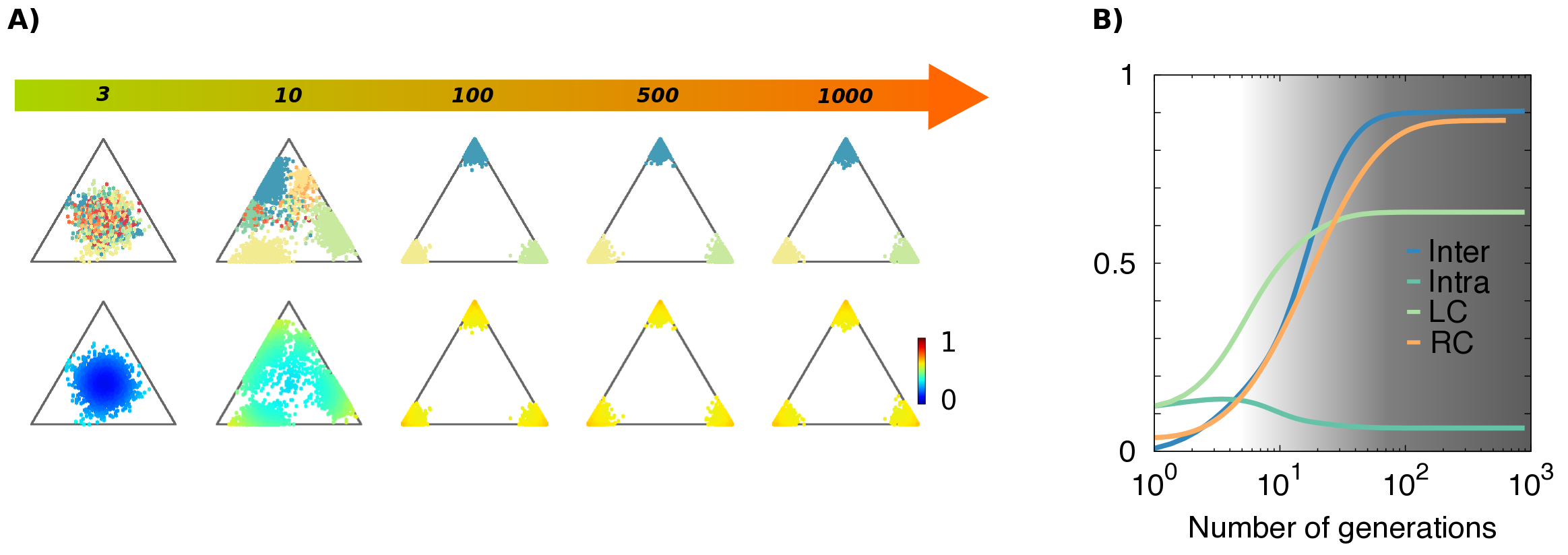}
 \caption{\textbf{(Color online) Evolution of the community with
     long-distance dispersal and global competition (i.e. well-mixed
     or mean-field dynamics).} A) Tradeoff space and complementarity
   measured at different generations. B) Inter and intra-specific
   distances, and local complementarity and relative complementarity 
   (RC= GC$_\mathrm{inter}$-GC$_\mathrm{intra}$) in time.  Parameters:
   $L=64$, $S=16$, $\beta=10$, $\mu=0.025$.}
 \label{fig:Appendix_mean-field}
\end{figure}

\subsection*{S4. Comparison with neutral theory ($\beta=0$)}
\label{Neutral system}

In the limit of no competition, $\beta=0$, our model equates to
neutral-theory \citep{Hubbell-Book} in which reproduction probabilities
become independent of individual phenotypes.
Fig. \ref{fig:Appendix_Neutral} reports computational results for this
case, illustrating the emergence of a very different scenario with
respect to the non-neutral case. Sexual reproduction still pulls
species together so they aggregate in the trade-off space, but their
centroids describe slow and independent random walks instead of being
controlled by a relatively fast separating drift (see
Fig. \ref{fig:Appendix_Neutral_2}).  This phenomenology is caused by
the lack of an effective force pushing species away; indeed segregated
species can become closer after some generations, but on average there
is only random drift allowing them to slowly diversify, so they cannot
account for the empirical observations in \citep{Zuppinger}.
Similarly, relative complementarities (which in the absence of
competition can be regarded as the averaged difference in the level of
phenotypic similarity between randomly sampled non-conspecific and
conspecific individuals) start to grow later and reach low values,

\begin{figure}[h!]
\centering
 \includegraphics[width=\textwidth]{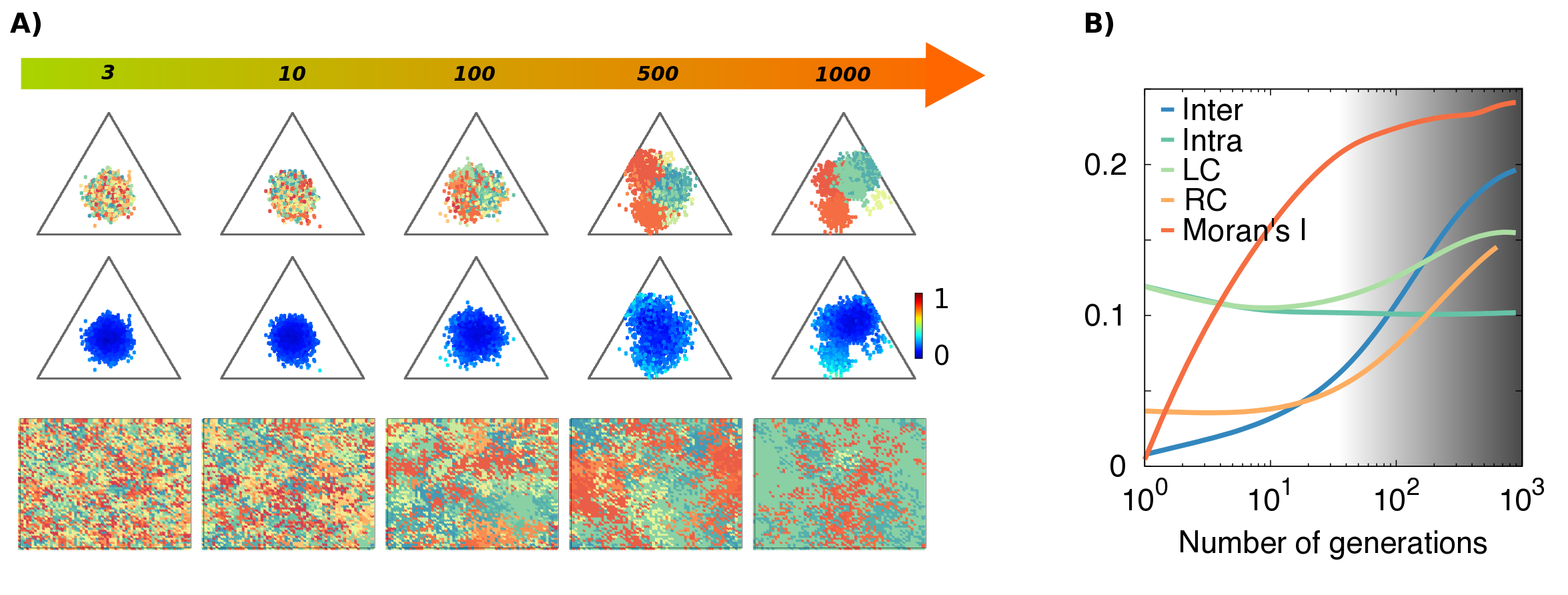}
 \caption{\textbf{(Color online) Neutral dynamics ($\beta=0$)},
   implying that all individuals have the same probability of
   reproduction independently of their species assignment and
   phenotype. A) Plots in the trade-off space illustrate that species
   hardly segregate in short time scales. In the physical space, local
   dispersal leads the the system to be clustered, i.e. positively
   auto-correlated rather than anti-correlated. B) Different measures
   illustrate that rapid evolutionary changes are much harder to
   observe in the neutral scenario. In particular, local
   complementarity (LC) and relative complementarity increase at a much
   slower pace.  The increase in the Moran's index confirms that the
   system remains positively correlated (as usually is the case in
   neutral models) Parameters have been set to
   $L=64$, $S=16$ and $\mu=0.025$.}
   \label{fig:Appendix_Neutral}
\end{figure}

\begin{figure}[h!]
\centering
 \includegraphics[width=0.4\textwidth]{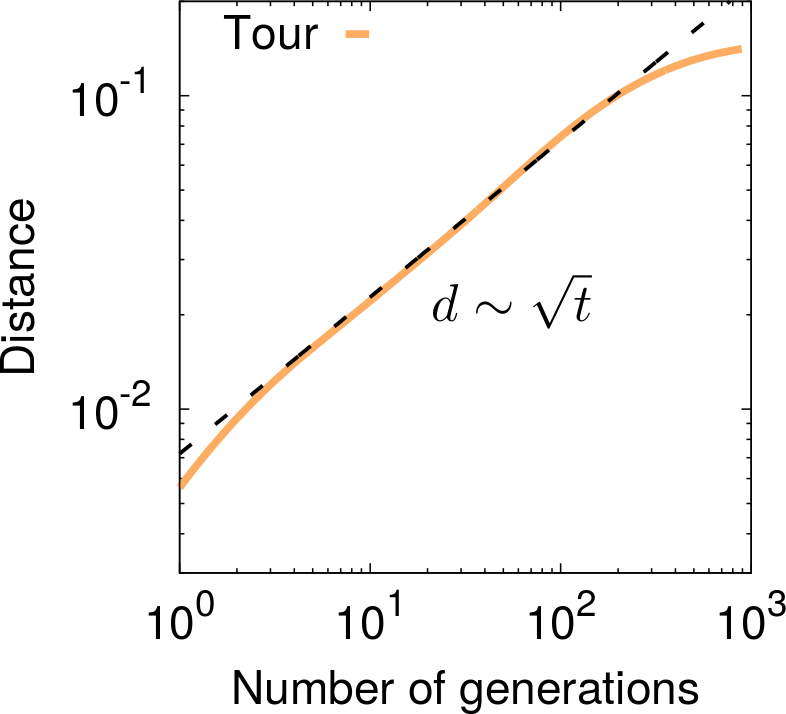}
 \caption{\textbf{(Color online) Species diffuse in the trade-off
     space under neutral dynamics ($\beta=0$):} The plot shows 
   average distance between species centroids and the central point of the
   phenotypic space (i.e. the position of all species centroids at $t=0$) as
   a function of time. Mutations cause a random movement of species centroids
   in the phenotypic space, as shown by the $0.5$ slope in double logarithmic scale,
   characteristic of diffusive processes. Parameter values are set as in
   Fig. \ref{fig:Appendix_Neutral_2}.}
 \label{fig:Appendix_Neutral_2}
\end{figure}

\subsection*{S5. Initial phenotypic traits}
\label{Initial conditions}
In the main text, initial conditions are given by randomly sampling
the value of each individual phenotypic trait from a single
distribution (Gaussian around the center of the phenotypic space),
independently of species labels. After some generations, we observe
that competition causes species to segregate in phenotypic space.

Although the most widely accepted definitions of species are based exclusively on the
role of mating barriers, individuals belonging to the same species tend to share 
common trait values. In this appendix, we approximate this kind of scenario and
test the robustness of our results running simulations with partial clustering of 
species in phenotypic space. For this, we sampled individual 
traits from equal amplitude Gaussian distributions centered around different (randomly
chosen) species-dependent points of the phenotypic space (see initial top panel in
Fig. \ref{fig:Appendix_initial_conditions}).

As illustrated in Fig. \ref{fig:Appendix_initial_conditions}, species
diversify sooner in phenotypic space (initial values for inter- and
intra-specific distances are higher than in the case described in the main
text) but after this transient difference asymptotic results remain
essentially unchanged. 

\begin{figure}[h!]
\centering
 \includegraphics[width=\textwidth]{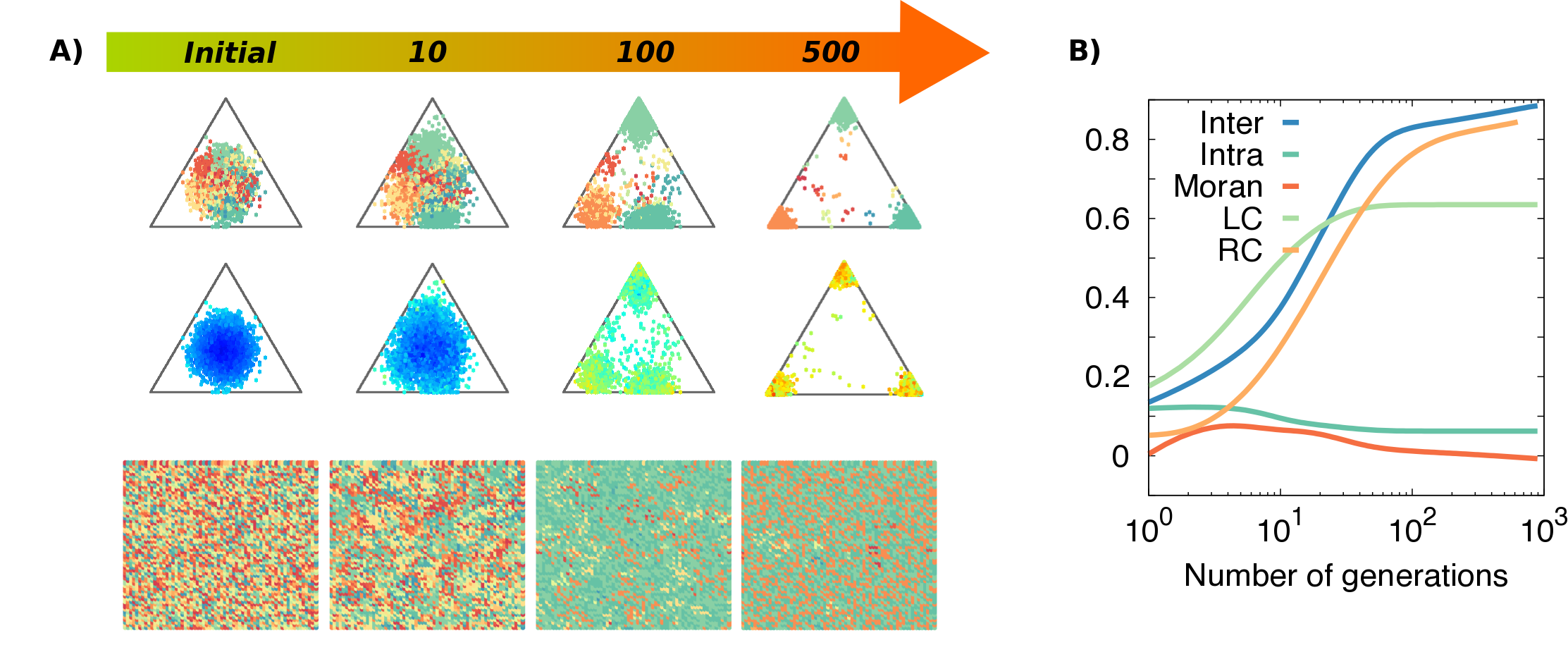}
 \caption{\textbf{(Color online) Simulations under initial phenotypic
 		segregation of species}. Individual traits are initially
   sampled from different species-dependent Gaussian
   distributions. Each of these Gaussians have a standard deviation
   equal to 0.05 and a mean value randomly selected from another
   Gaussian (with the same amplitude 0.05) centered at the triangle
   barycenter.  A) From top to bottom: Tradeoff values, trait
   complementarities, and spatial distributions at different
   generations. B) Evolution of the inter and intra-specific
   distances, local and relative complementarity and Moran's Index.
   Parameters: $L=64, S=16,\beta=10, \mu=0.025$.}
\label{fig:Appendix_initial_conditions}
\end{figure}

\subsection*{S6. Effect of the competition kernel}
\label{Competition kernels}

In the implementation of our model presented in the main text, the
reproduction probability of an individual $i$ is proportional to
$e^{\beta C_i}$, where $C_i$ is the average trait complementarity among neighbors,
$C_i=\frac{1}{K}\sum_ j \frac{1}{n}\sum^k |T^k(i)-T^k(j)|$. However,
the use of a non-differentiable argument (absolute value) appearing
linearly in the exponential kernel may lead to spurious robust
coexistence of arbitrarily similar species (at zero phenotypic
distance) \citep{Geritz1999,
  Adler-Mosquera2000,Hernandez2009,Barabas2013,Lampert2014}.  
Kernels of the form $e^{\beta C_i^{2+\alpha}}$ with $\alpha>0$ have been shown to avoid such artifacts
\citep{Pigolotti2007,Pigolotti2010}. For these reasons, we also
considered an alternative competition kernel of the form
$e^{\beta C_i^4}$ to check for the validity and robustness of our
conclusions.

Results are shown in Fig \ref{fig:Appendix_kernels}. We observe that,
as the quartic kernel reduces the overall competition of phenotypically similar
individuals, it leads to a slower species-diversification process (as
compared with the linear one for the same value of the
parameters). However, results are qualitatively similar to the linear
kernel case. In particular, similar (equivalent) species continue to
emerge and coexiste for very long times, as in the linear case.  In
Appendix S\ref{sec:appendix-neutral} we discuss the coexistence of
emergent equivalent species in more detail.

\begin{figure}[h!]
\centering
 \includegraphics[width=\textwidth]{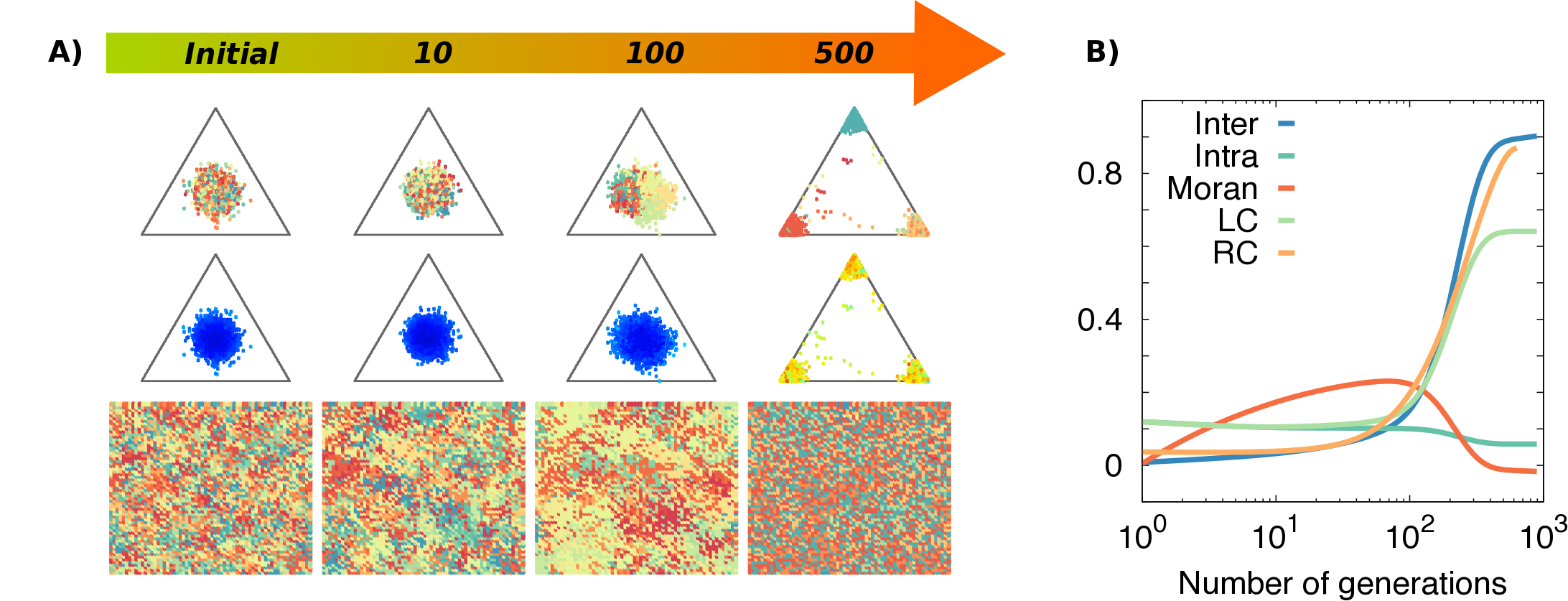}
 \caption{\textbf{(Color online) Quartic competition kernel}.  A) From
   top to bottom: Tradeoff values, trait complementarities, and
   spatial distributions at different generations. Simulations were
   run setting the performance of individuals proportional to
   $e^{\beta C^4}$ (rather than $e^{\beta C}$ used in the main text),
   where $C$ is the average trait complementarity among neighbors. B)
   Inter and intra-specific distances, local and relative complementarity and
   Moran's Index evolution.  Parameters:
   $L=64, S=16,\beta=10, \mu=0.025$.}
\label{fig:Appendix_kernels}
\end{figure}

\subsection*{S7. Emergent Neutrality}
\label{sec:appendix-neutral}
In the main text, we discuss the possibility of emerging
  phenotypically-equivalent species coexisting for long periods of
  time. It is important to underline that not all realizations lead to
  equivalent species coexisting in the community. However, this
  appears to be a significant pattern and we explored it further. In
  particular, we decided to check the stability of the coexistence.
  In this Section, we study the mean coexistence time of equivalent
  species as a measure of coexistence stability.

  Two equivalent species coexist until one of them invades the
  phenotypic space of the other as a result of demographic
  fluctuations. As a first step to quantify the dynamics of this
  process, we define a computational criterion to determine
  equivalence: two species $s_1$ and $s_2$ are considered equivalent
  if their inter-specific distance (i.e. distance between their
  centroids) differs less than a fraction of their mean intraspecific
  distance (mean trait amplitude), for instance $1/4$, which produces
  a significant overlap of the clusters of both species in phenotypic
  space.

  We then measured the mean number of generations $\Delta T$ between
  the time at which $4$ species remain in the system (with two of them
  being equivalent, based on the previous definition) and the time at
  which one of such equivalent species invades the other one. In voter
  models (i.e. the neutral case), the mean time to reach
  mono-dominance, $\Delta T$, increases with the number of individuals
  in the community, $N$; in particular, $\Delta T\sim N\log N$ in a 2D
  lattice and $\Delta T\sim N$ in a well-mixed situation (for
  instance, the case of long-distance dispersal and global
  competition) \citep{Cox1989}.

  Fig. \ref{fig:Appendix_consensus} shows $\Delta T$ for simulations
  with limited dispersal (i.e. the 2D case), as well as the
  theoretical expectation for the neutral case. To check the
  robustness of coexistence to the shape of the competition kernel
  \citep{Geritz1999,
    Adler-Mosquera2000,Pigolotti2007,Hernandez2009,Pigolotti2010,Barabas2013,Lampert2014}
  (see Appendix S\ref{Competition kernels}), we run simulations using a
  linear-exponential ($e^{\beta C}$) and a quartic-exponential
  ($e^{\beta C^4}$) kernel, where $C$ is the mean trait
  complementarity among neighbors. In both cases, our results are
  compatible with the neutral scenario.

\begin{figure}[h!]
\centering
 \includegraphics[width=0.5\textwidth]{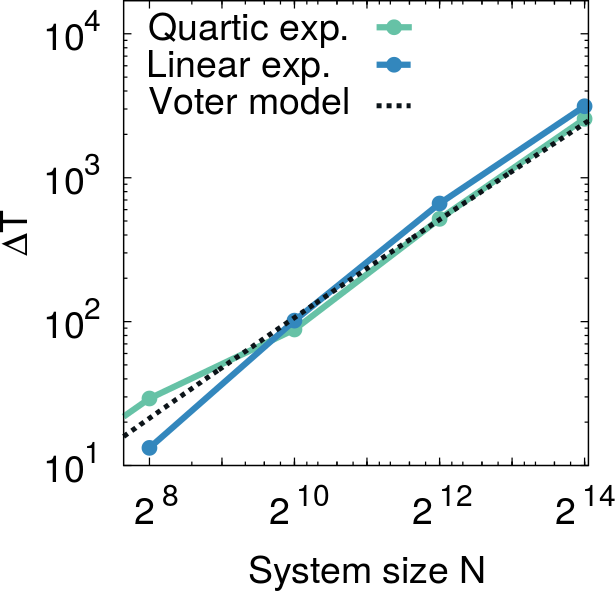}
 \caption{\textbf{(Color online) Mean number of \lq\lq{}
     coexistence\rq\rq{}}, for different system sizes of the
   community. We show results using a linear-exponential
   ($e^{\beta C}$) and a quartic-exponential ($e^{\beta C^4}$) kernel
   comparing them with the theoretical expectation for a 2D voter
   model, $\Delta T\propto N\log N$ \citep{Cox1989}. Parameters:
   $L=64, S=8,\beta=10, \mu=0.025$. Deviations from the straight line
   probably stem from lack of statistics (which is costly at such
   large sizes/times). }
\label{fig:Appendix_consensus}
\end{figure}

\subsection*{S8. Asymmetrical resource trade-offs}
\label{Symmetry breaking}

In this section we consider a model variant in which positions in the
trade-off space are \emph{not} equally rewarding a priori. In
particular, we chose one of the corners to be favored respect to the
others: individuals whose phenotypes are closer to that vertex have a
higher probability of reproduction. This could be interpreted as
  one particular limiting resource being more crucial, or mean that
  the availability of some resource scales nonlinearly with
  corresponding trait (e.g., a plant with a short root might not be
  able to reach a deep water layer.  Individuals with longer root
  systems will have a disproportionate advantage).

In particular, we now modulate the performance of each individual $i$
by multiplying it by a factor $R_i=r_1T^1+r_2T^2+r_3T^3$; where $r_1$,
$r_2$ and $r_3$ are weights (real numbers in the interval $[0,1]$)
such that $r_1+r_2+r_3=1$. For simplicity we fix
$r_1=(1+2\epsilon)/3$, $r_2=r_3=(1-\epsilon)/3$ in order to control
the asymmetry with a single parameter ($\epsilon$; the
symmetric case is $\epsilon=0$), while $\epsilon$
values close to $1$ lead to large asymmetries.  In what follows we 
fix $\epsilon=0.99$.

As illustrated in Fig. \ref{fig:Appendix_Biased}, although most of
individuals initially occupy the most favored (left) corner, after a
few generations, some individuals also settle at other available (and
less favorable) regions; this is a consequence of the system's tendency
to reduce the level of competition. In conclusion, the main
phenomenology reported in the main text appears robust to asymmetrical trade-offs.

\begin{figure}[h!]
\centering
 \includegraphics[width=\textwidth]{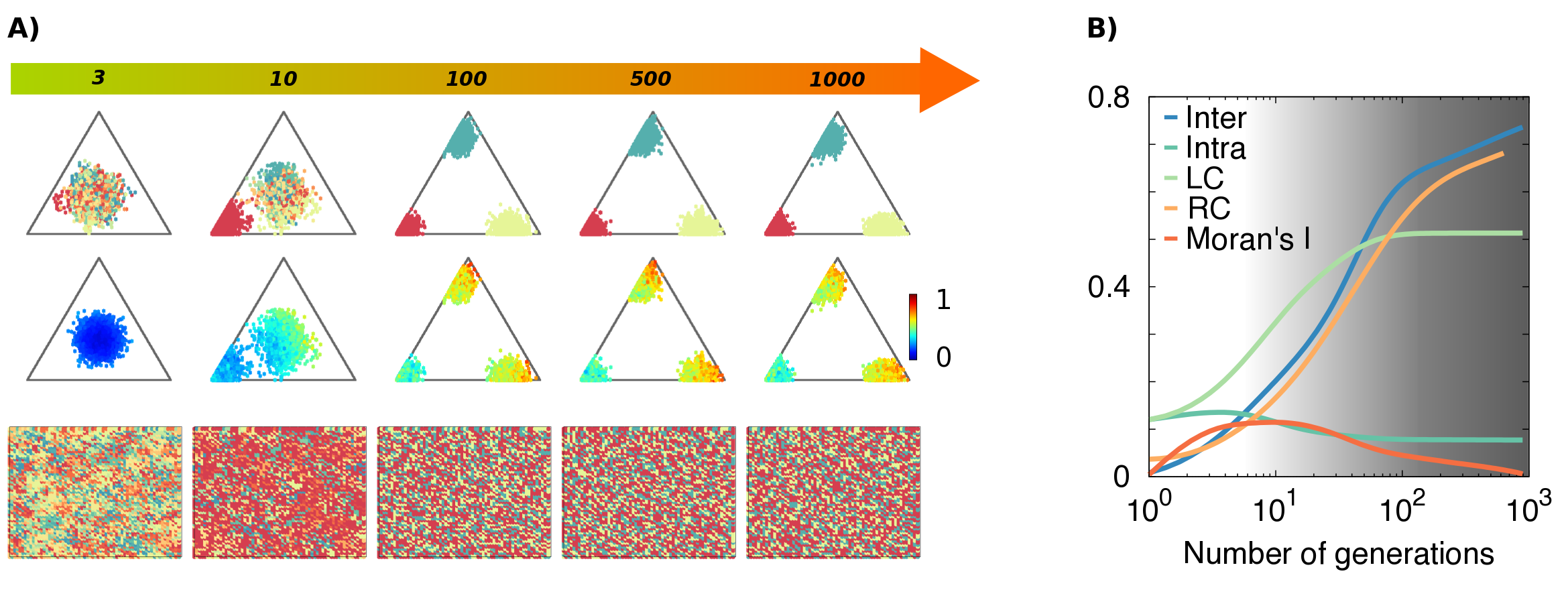}
 \caption{\textbf{(Color online) Asymmetry among trade-offs}: A)
   Trade-off space and complementarity as a function of the number of
   generations. B) Different measurements characterizing the community
   in time. Local and relative complementarities confirm that rapid
   evolutionary changes may within a few generations.  Parameters:
   $L=64$, $S=16$, $\beta=10$, $\mu=0.025$.}
   \label{fig:Appendix_Biased}
\end{figure}

\subsection*{S9. Asexual reproduction}
\label{Asexual reproduction}
Our model adopts the (sexual) reproduction mechanism of the
communities considered in the experiments by Zuppinger-Dingley \emph{et al.}
\citep{Zuppinger}. Here we analyze a case of asexual reproduction in
which the traits are directly transmitted from an individual to its
offspring (with some variability), i.e. taking $\eta=1$ in our model
(see main text).

Fig \ref{fig:Appendix_Asx}A shows the evolution of individual
phenotypes (each trait value $T_1,T_2,T_3$ is represented by the amount
of red, yellow and blue respectively), complementarity and spatial
distribution.  Observe that, once again, the chief phenomenology of
the model, i.e. segregation toward high levels of specialization, is
observed. However, in this case, the mechanism of
  diversification is quite different: individuals from any given
  species can specialize independently. Thus, in the absence of
  sexual reproduction, diversification occurs at an individual rather
  than at a species level. This type of individual differentiation
  fosters the presence of equivalent species (as the species label
  becomes completely irrelevant in this setting).

\begin{figure}[h!]
\centering
 \includegraphics[width=\textwidth]{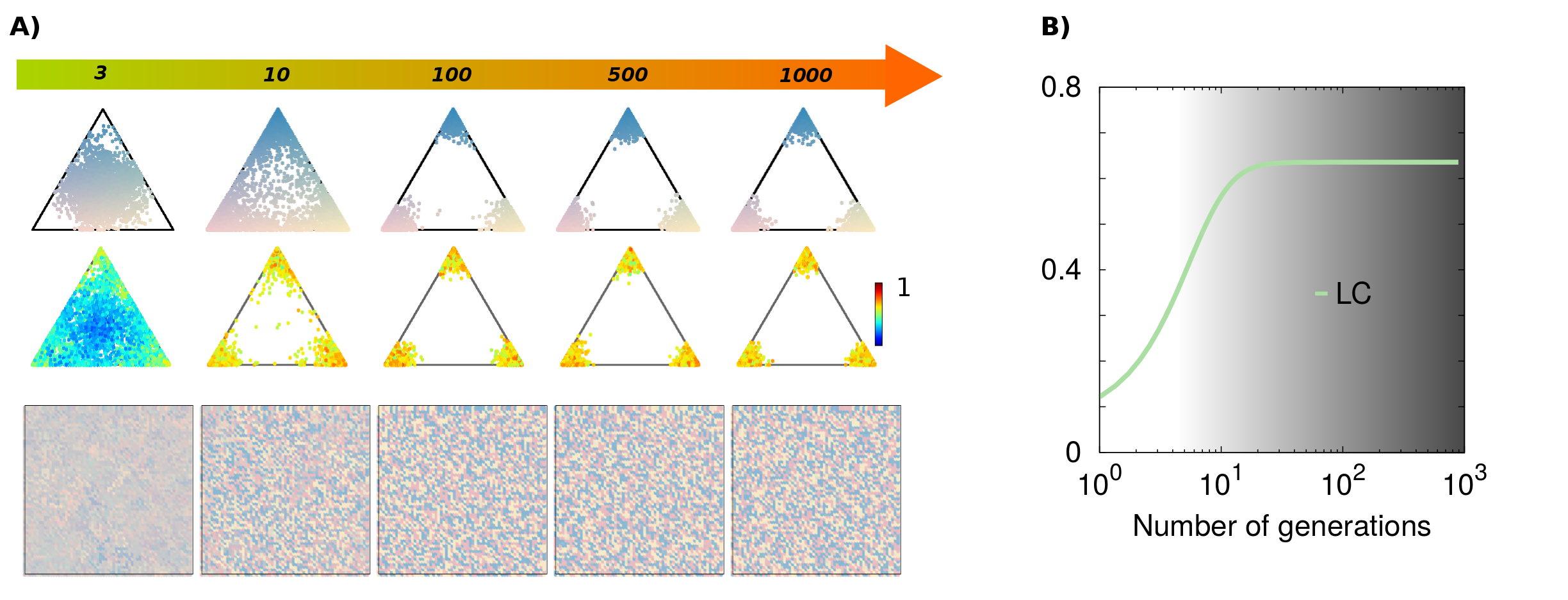}
 \caption{\textbf{(Color online) Asexual reproduction}: A) Tradeoff
   space and complementarity as a function of the number of
   generations. Each individual traits values $T_1,T_2,T_3$ are
   represented by the amount of red, yellow and blue
   respectively. Parameter values: $L=64$, $S=16$, $\beta=10$,
   $\mu=0.025$.  Competition avoidance leads individuals to segregate
   in the trade-off space. B) Local complementarity increases
   similarly to the main model.}
   \label{fig:Appendix_Asx}
\end{figure}

\subsection*{S10. Surviving species}
\label{Surviving species}

As we do not include mechanisms such as migration or speciation, the
number of species actually present in the community can be reduced
after several generations. The resulting change in diversity can be
regarded as an important attribute, because it illustrates the limit of maximum
diversity that a finite system can harbor is the absence of
inmigration or speciation processes.

Fig. \ref{fig:Appendix_S3} shows the number of surviving species in
different scenarios, including different levels of competition and
variability parameters, and other model variants. As expected,
extinctions occur more rapidly for higher levels of competition
(larger values of $\beta$). The effective level of competition is
enhanced in the mean-field case in which all individuals interact with
each other (see Appendix S\ref{mean-field}), leading to faster
extinction.

In the neutral case ($\beta=0$), species disappear at a very slow rate
as there is no competition, but due to stochasticity, most of them are
likely to disappear, leading to mono-dominance for sufficiently large
timescales \citep{Liggett}. Interestingly, the stable solution in our
model with competition ($\beta \neq 0$) consists of multiple species
--as many as the niche dimensionality, in this case 3-- coexisting for
an arbitrarily large number generations. This result is congruent with
the ``niche dimension hypothesis'', which states that a greater
diversity of niches leads to a greater diversity of species
\citep{Hutchinson}.

\begin{figure}[h!]
\centering
 \includegraphics[width=\textwidth]{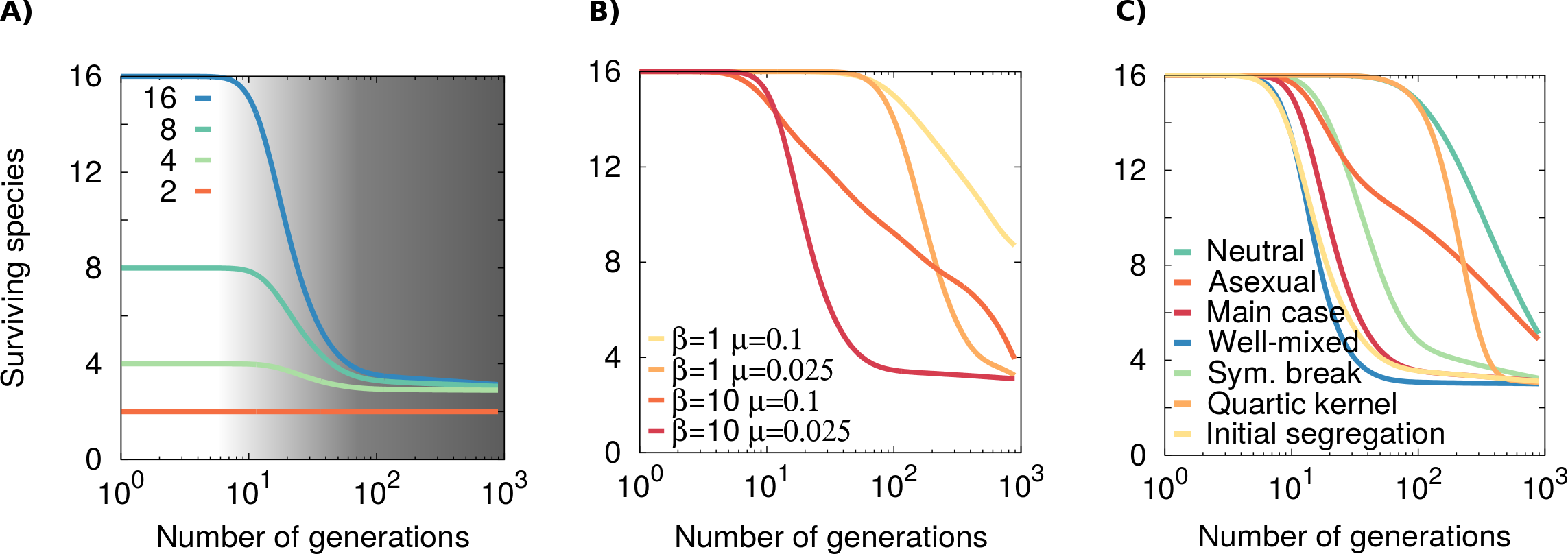}
 \caption{\textbf{(Color online) Number of surviving species in time
     for different A) initial number of species, B) competition and
     variability parameters and C) variants of the model}. Species
   disappear faster in environments with high competition (higher values of
   $\beta$, or the mean-field). In contrast, the neutral case
   ($\beta=0$) corresponds to the case in which more species survive
   after generations, although, due to demographic fluctuations, they
   still disappear on the long term. For sufficiently large numbers of
   generations, the system converges to a state with the same number
   of species than the niche dimensionality ($3$ in our case); these
   species coexist for arbitrarily long periods (provided the lattice
   is sufficiently large). Parameter values: $L=64$, and $\beta=10$,
   and $\mu=0.025$ in A) and C).}
\label{fig:Appendix_S3}
   \end{figure}

\bibliographystyle{plainnat}
\bibliography{biblio-plos.bib}

\end{document}